\documentclass[twoside,twocolumn,tightenlines,superscriptaddress,floatfix,showpacs,aps,prl]{revtex4-2}
\usepackage{graphicx}
\usepackage{wasysym}
\usepackage{amssymb}
\usepackage{bbold}
\usepackage[makeroom]{cancel}
\usepackage[utf8]{inputenc}
\usepackage[normalem]{ulem}
\newcommand{\be}{\begin{equation}}
\newcommand{\ee}{\end{equation}}
\usepackage[unicode=true,pdfusetitle,
 bookmarks=true,bookmarksnumbered=false,bookmarksopen=false,
 breaklinks=false,pdfborder={0 0 1},backref=false,colorlinks=false]
 {hyperref}
\hypersetup{
 colorlinks,linkcolor=blue,citecolor=blue,urlcolor=blue}

\begin{document}

\preprint{APS/123-QED}

\title{Extrinsic Orbital Hall Effect and Orbital Relaxation in Mesoscopic Devices}

\author{Anderson L. R. Barbosa}
\email{anderson.barbosa@ufrpe.br}
\affiliation{Departamento de F\'{\i}sica, Universidade Federal Rural de Pernambuco, 52171-900, Recife, PE, Brazil}
\author{Hyun-Woo Lee}
\affiliation{Department of Physics, Pohang University of Science and Technology (POSTECH), Pohang 37673, Korea}
\author{Tatiana G. Rappoport}
\email{tgrappoport@fisica.uminho.pt}
\affiliation{Centro Brasileiro de Pesquisas Físicas , R. Dr. Xavier Sigaud, 150, 4715-330 Rio de Janeiro, RJ, Brazil}
\affiliation{Physics Center of Minho and Porto Universities (CF-UM-UP),Campus of Gualtar, 4710-057, Braga, Portugal}
\affiliation{International Iberian Nanotechnology Laboratory (INL), Av. Mestre José Veiga, 4715-330 Braga, Portugal}

\date{\today}

\begin{abstract}
Despite recent advances in orbitronics, the influence of disorder on the orbital Hall effect and orbital relaxation mechanisms remains poorly understood. In this work, we numerically investigate the role of disorder in orbital transport within mesoscopic devices using a real-space tight-binding model on a two-dimensional square lattice that hosts atomic orbitals capable of carrying atomic orbital angular momentum. By considering devices with varying geometries—square and rectangular—and systematically tuning disorder strength, we examine the disorder effect on orbital Hall current (OHC)  generation, and orbital relaxation. Our results reveal a strong dependence of the OHC and orbital Hall angle on disorder strength. In square devices, we demonstrate that the orbital Hall response can be strongly enhanced by disorder and its dependence on the disorder strength indicates the dominance of skew-scattering mechanism in the diffusive regime. In rectangular geometries, the orbital current decays exponentially with increasing device width, from which the orbital relaxation length is extracted. These findings provide critical insights into disorder-driven orbital transport phenomena and lay the foundation for designing next-generation orbitronic devices.

\end{abstract}


\maketitle


The manipulation of orbital angular momentum (OAM) in solids has emerged as a frontier in condensed matter physics. Orbitronics offers a versatile platform for both fundamental discoveries and technological applications~\cite{Go-Review, Choi2023, Lyalin2023, Sala2023,Ding2020, Sala2022,Jo2024,Lee2021,Ding2024,Han2022,Busch2024,Seifert2023,Kumar2023,Xu2024,Go2023,Ning2025,Han2023b,Santos2023,PhysRevApplied.22.064071,PhysRevLett.134.026702,Hayashi2024,Johansson2021,ElHamdi2023,PhysRevLett.134.036304}, extending the reach of spintronics and enabling next-generation device concepts~\cite{Go-Review, Jo2024}. A central mechanism in this field is the orbital Hall effect (OHE), which generates transverse OAM currents in response to an electric field~\cite{Bernevig2005,Kontani2008, Go2018,Salemi2022,Canonico2020,Cysne2021,Costa2023}, analogous to the spin Hall effect. Only recently observed in light metals~\cite{Choi2023, Lyalin2023, Sala2023}, the OHE opens new directions for exploring orbital transport.

Disorder, inherent to real devices, adds complexity to orbital transport. In spintronics, disorder-induced mechanisms like skew scattering and side jump are key extrinsic sources of the spin Hall effect~\cite{Sinova2015,Fert2011}. Similarly, in orbitronics, disorder may play an active role in generating OHE and could be used for device engineering. Despite its relevance, the impact of disorder on the OHE remains poorly understood, with only a few theoretical works addressing it~\cite{Pezo2023,Liu2023, Fonseca2023,Barbosa2024, Tang2024, Canonico2024,Veneri2025}.

Theoretical studies of extrinsic mechanisms have mainly addressed the itinerant contribution to OAM in systems where conduction electrons stem from OAM-inactive orbitals~\cite{Pezo2023,Liu2023,Canonico2024,Veneri2025}. Results remain conflicting. While perturbative approaches suggest dominance of an extrinsic mechanism insensitive to disorder concentration~\cite{Liu2023}, other works report suppression of the intrinsic OHE by short-range disorder~\cite{Tang2024}. Real-space simulations reveal disorder-dependent OHE~\cite{Canonico2024}, supported by recent non-perturbative results identifying skew scattering as the leading mechanism in the diffusive regime~\cite{Veneri2025}.

Another key factor in orbital transport efficiency is orbital relaxation—the decay of OAM currents due to disorder-induced scattering. It is characterized by the orbital relaxation length $\lambda_L$ or relaxation time, which describe the spatial or temporal decay of orbital polarization, analogous to spin relaxation in spintronics.

Experimental measurements of the orbital relaxation length $\lambda_L$ vary widely across materials and methods. In titanium, values of 50–60~nm~\cite{Choi2023} and 47~nm~\cite{Sala2023} have been reported, while in chromium, significantly shorter lengths of 6–7~nm have been observed~\cite{Lyalin2023}. Theoretical studies on orbital relaxation have only recently emerged~\cite{Sohn2024,Rang2024,Kabanov2024}, proposing different mechanisms: from slow decay dominated by Dyakonov–Perel-like processes~\cite{Sohn2024,Kabanov2024}, to shorter relaxation lengths~\cite{Rang2024}, and possible Elliott–Yafet-type behavior~\cite{Sohn2024,Kabanov2024}.

The role of disorder in shaping the OHE and orbital relaxation remains poorly understood. To address this, we take a different approach: instead of focusing on bulk systems as in most prior studies, we examine mesoscopic devices. These provide a distinct platform to probe disorder effects and gain insight into orbital relaxation and disorder-induced scattering~\cite{Fonseca2023,Barbosa2024}.

We explicitly consider the atomic orbital degree of freedom, focusing on OAM from atomic orbitals rather than from itinerant circulation. Through numerical simulations, we investigate how disorder influences both extrinsic enhancements of the OHE and the suppression of orbital coherence.

We consider a centrosymmetric 2D square lattice modeled by a nearest-neighbor tight-binding Hamiltonian \cite{Go2018}, with four orbitals per site (one $s$ and three $p$ orbitals) and lattice constant $a$. The $p$ orbitals carry atomic OAM, making the system orbital active. The  Hamiltonian is given by \cite{Sahu2021}
\begin{eqnarray}
 H&=&\sum_{\langle i,j\rangle\alpha\beta\sigma}t_{i\alpha,j\beta}c_{i\alpha\sigma}^{\dagger} c_{j\beta\sigma}+ \sum_{i\alpha\sigma} E_{i\alpha\sigma} c_{i\alpha\sigma}^\dagger c_{i\alpha\sigma}\nonumber\\&+& \lambda_{\rm SOC} \sum_{i\alpha\beta\sigma\delta}\sum_\gamma c_{i\alpha\sigma}^{\dagger}L^\gamma_{\alpha\beta}S^\gamma_{\sigma\delta}c_{i\beta\delta}, \label{TBH}
\end{eqnarray}
Here, ${i,j}$, ${\alpha,\beta}$, and ${\sigma,\delta}$ index unit cells, orbitals, and spins, respectively, with $\gamma = {x, y, z}$. The first term represents nearest-neighbor hopping, where $c_{i\alpha\sigma}$ ($c_{i\alpha\sigma}^\dagger$) are annihilation (creation) operators, and $t_{i\alpha,j\beta}$ are hopping integrals. The second term is the on-site energy $E_{i\alpha\sigma}$, including the orbital energy $E_{\alpha\sigma}$ and an Anderson disorder term $\epsilon_{i\alpha\sigma}$. We assume disorder is orbital- and spin-independent, implemented via an electrostatic potential $\epsilon_i$ randomly varying per site, drawn from a uniform distribution in $\left(-U/2,U/2\right)$, with $U$ the disorder strength. The last term is spin-orbit coupling (SOC), with strength $\lambda_{\rm SOC}$, angular momentum $\vec{L}$, and spin-$1/2$ operator $\vec{S}$.

We use typical Hamiltonian parameters (in eV): on-site energies $E_s = 3.2$, $E_{p_x} = E_{p_y} = E_{p_z} = -0.5$; and nearest-neighbor hoppings $t_s = 0.5$, $t_{p\sigma} = 0.5$, $t_{p\pi} = 0.2$. The $sp$ hopping $t_{sp}$ mediates $k$-dependent hybridization among $p_x$, $p_y$, and $p_z$ orbitals. When $t_{sp} = 0$, the orbital texture vanishes, and the intrinsic OHE mechanism is suppressed \cite{Go2018}.

\begin{figure}
\includegraphics[width=.45\textwidth]{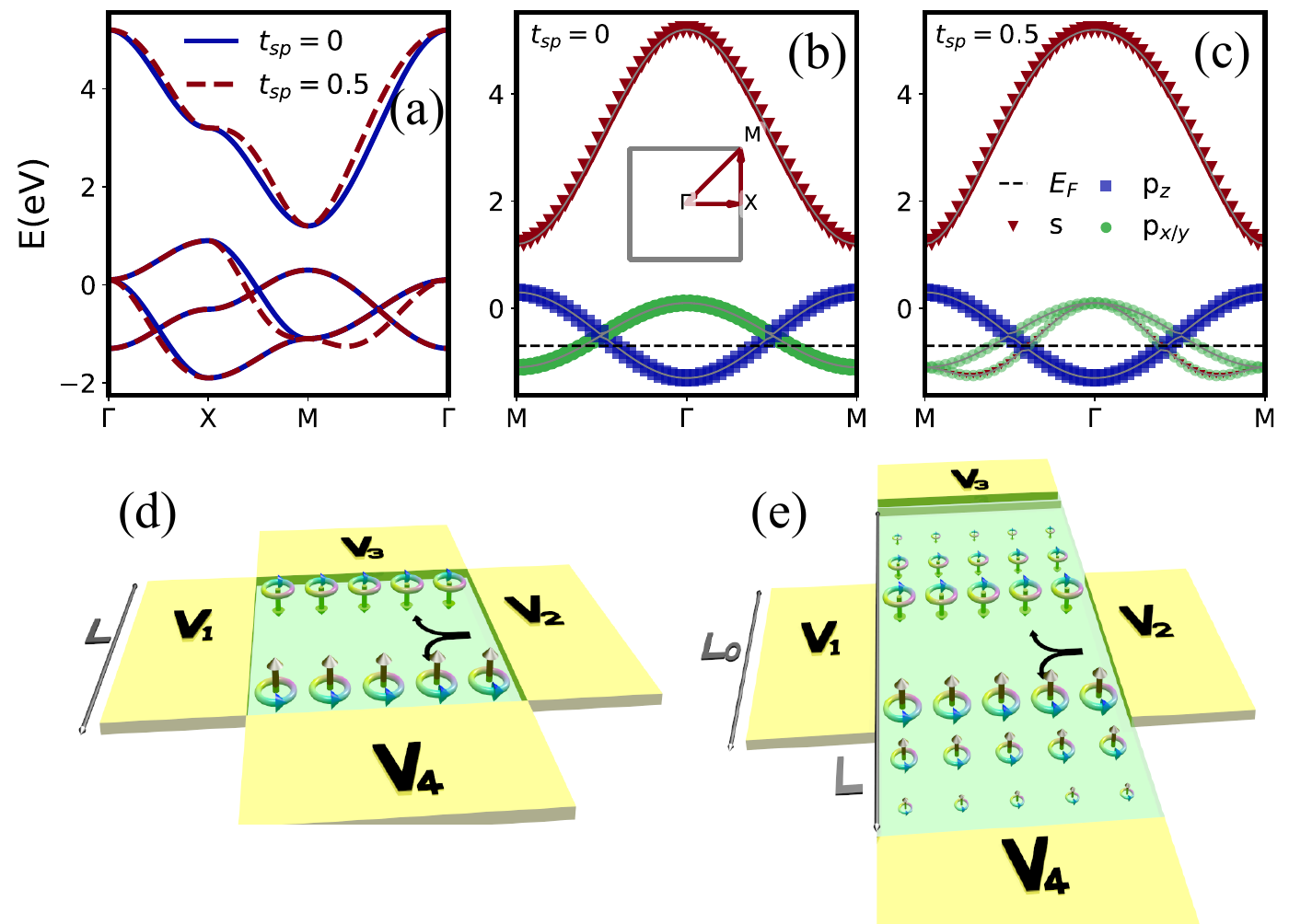}
\caption{(a) Bulk energy bands of a 2D square lattice with four orbitals per site for $t_{sp} = 0$ (blue) and $t_{sp} = 0.5$ eV (dashed red), along the Brillouin zone path shown in (b). (b)-(c) Energy bands and projected orbital characters along M–$\Gamma$–M. The upper band has $s$ character; the lower bands are mainly $p$-like, with two degenerate at $t_{sp}=0$ (lowest band). (b) $t_{sp}=0$; (c) $t_{sp}=0.5$ eV. Dashed lines indicate the Fermi energy $E_F = - 0.7$ eV. (d)-(e) OHE in devices where the orbitally active region is a square lattice with momentum-space orbital texture, connected to four semi-infinite terminals under potentials $V_1 = V/2$ and $V_2 = -V/2$. (d) Square device of size $L \times L$. (e) Rectangular device of size $L_0 \times L$, with transverse orbital current flowing across width $L$.}\label{fig01}
\end{figure}

Figure~\ref{fig01}(a) shows bulk energy bands of the 2D square lattice for different $t_{sp}$ values, along a Brillouin zone (BZ) path indicated in the inset of Fig.\ref{fig01}(b). At $t_{sp} = 0$, two bands are degenerate along $\Gamma$–M, forming degenerate lines in the BZ. Figures\ref{fig01}(b–c) show the energy bands along one such line, with projected orbital character represented by symbol size for $s$, $p_{x/y}$, and $p_z$ orbitals. The upper band is mostly $s$-like; the lower bands are $p$-dominated. Because the system is strictly 2D, $t_{sp}$ does not couple to $p_z$ orbitals and the $p_z$ band remains unaffected. For $t_{sp} = 0$ [Fig.\ref{fig01}(b)], no OHE occurs. For finite $t_{sp}$ [Fig.\ref{fig01}(c)], hybridization among $s$, $p_x$, and $p_y$ lifts the degeneracy. The resulting $p_{x/y}$ states near the Fermi level drive the intrinsic OHE.

Our OHE setup consists of two mesoscopic devices, each with a central region governed by a Hamiltonian including both orbital and spin degrees of freedom. As shown in Fig.~\ref{fig01}, both connect to four semi-infinite terminals under voltages $V_i$. The first device [Fig.~\ref{fig01}(d)] is square, of size $L \times L$; the second [Fig.~\ref{fig01}(e)] is rectangular, with fixed length $L_0$ and variable width $L$. Simulations were performed using the KWANT package~\cite{Groth2014}. A fixed longitudinal bias is applied with $V_1 = V/2$ and $V_2 = -V/2$, driving a charge current from terminal 1 to 2 and inducing an orbital Hall current (OHC) across terminals 3 and 4. In devices with strong spin-orbit coupling, a spin Hall current (SHC) also emerges, transverse to the charge flow.

From the Landauer-B\"uttiker formalism, the orbital (spin) projected current through the $i$th terminal in the linear regime at low temperature is given by $I^{\mathcal{O}}_{i,\eta} = \frac{e^2}{h}\sum_{j} \tau_{ij,\eta}^{\mathcal{O}} \left( V_i - V_j \right)$
 with $\mathcal{O}$ being one of the $L_z$ orbital angular momentum ($\circlearrowleft$ or $\circlearrowright$)  or $S_z$ spin states ($\uparrow$ or $\downarrow$).
The orbital (spin) transmission coefficient is calculated from the scattering matrix $\mathcal{S}=\left[\mathcal{S}_{ij}\right]_{i,j=1,...,4}$ as
\begin{equation}
\tau_{ij,\eta}^{\mathcal{O}} =\textbf{Tr}\left[\left(\mathcal{S}_{ij}\right)^{\dagger} \mathcal{P}^{\mathcal{O}}_{\eta} \mathcal{S}_{ij}\right], 
\end{equation}

\noindent where the matrices $\displaystyle \mathcal{P}^{L_\eta}_{\eta} =  l^\eta \otimes \sigma^0$ and $\mathcal{P}^{S_\eta}_{\eta} = l^0 \otimes \sigma^\eta$ are projectors. The matrices $l^{\eta}$ and $\sigma^{\eta}$ with $\eta=\lbrace x,y,z\rbrace$ are the orbital angular momentum and spin matrices, respectively, and the cases with $\eta=0$ refer to the identity matrices in the orbital and spin subspaces. Thus, setting either $\eta=0$ or $\eta=\{x,y,z\}$, the charge, orbital and spin can be addressed. {The trace is carried over transport modes in the terminals $i$ and $j$, and $\mathcal{S}_{ii}$ and $\mathcal{S}_{ij}$ are  reflection and transmission block matrices of $\mathcal{S}$,  respectively.

\begin{figure}
\includegraphics[width=1\columnwidth]{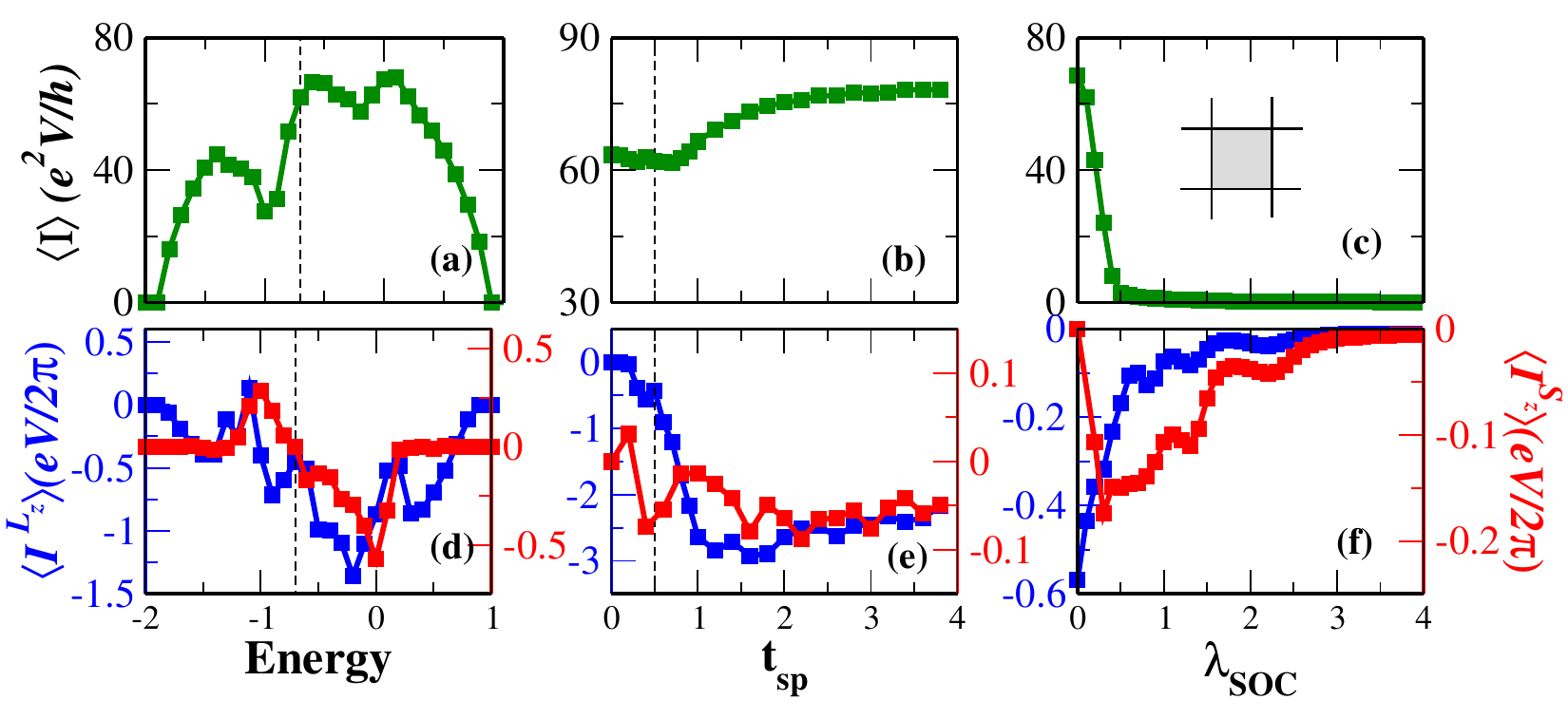}
\caption{(a-c) Longitudinal charge current, (d-f)  OHC  and SHC through a square device with dimension $80a \times 80a$ (depicted in Fig.\ref{fig01}.d), as a function of Fermi energy $E_F$ (left column), $sp$ hopping $t_{sp}$ (central column) and SOC strength $\lambda_{SOC}$ (right column). For all panels, the Anderson disorder strength is kept fixed at $U=0.5$ eV. For the left panels,  $t_{sp}=0.5$ eV and $\lambda_{\rm SOC} = 0.1$ eV; for the central panels, $E_F = -0.7$ eV and $\lambda_{\rm SOC} = 0.1$ eV; for the right panels, $E_F = -0.7$ eV and $t_{sp}=0.5$ eV. The dashed lines are (a,d) $E_F=-0.7$ eV (b,e) $t_{sp}=0.5$ eV, and (c,f) $\lambda_{\rm SOC}=0.1$ eV. The dashed lines in (c,f) overlap with the left vertical axes and are not visible. All current values are averaged over 200 realizations of disorder. The dependence of charge, orbital, and spin currents with disorder $U$ is discussed in the SM \cite{SM}.}\label{fig02}
\end{figure}

Pure OHC (SHC) $I^{L_z(S_z)}_{i,z}=\frac{\hbar}{e}(I^{\circlearrowleft(\uparrow)}_{i}-I^{\circlearrowright(\downarrow)}_{i})$, $i=3,4$ can be obtained assuming that the charge current vanishes at the transverse terminals, $I^{c}_{i,0}=I^{\circlearrowleft(\uparrow)}_i+I^{\circlearrowright(\downarrow)}_i=0$, while the charge current is conserved in the longitudinal terminals, $I^c_{1,0}=-I^c_{2,0}=I^c$. By applying these conditions to the orbital (spin) projected current, we obtain \cite{Fonseca2023,PhysRevB.72.075361,PhysRevLett.98.196601} 
$
I^{\mathcal{O}}_{i,\eta} = \frac{e}{2\pi}\left[\left(\tau_{i2,\eta}^{\mathcal{O}}-\tau_{i1,\eta}^{\mathcal{O}}\right)\frac{V}{2}
- \left(\tau_{i3,\eta}^{\mathcal{O}}V_3 + \tau_{i4,\eta}^{\mathcal{O}}V_4\right)\right]$, 
for $i=3,4$, where $V$ is a constant potential difference between the longitudinal terminals, and $V_{3,4}$ is the transversal terminal voltage.

In the device shown in Fig.~\ref{fig01}(d), the OHC is generated uniformly across the entire $L \times L$ area by a charge current flowing through the longitudinal terminals and is detected at the transverse ones. This configuration is ideal for characterizing OHC generation. In contrast, the device in Fig.~\ref{fig01}(e) generates the OHC within a smaller $L_0 \times L_0$ region, while detection occurs at transverse terminals separated by a much larger distance $L \gg L_0$. By varying $L$ while keeping $L_0$ fixed, one can probe orbital relaxation. This ability to decouple generation and detection regions is a key advantage of mesoscopic devices.

We begin with the device shown in Fig.~\ref{fig01}(d). Figure~\ref{fig02} presents the longitudinal charge current (panels a–c), along with the OHC and SHC (panels d–f), for square devices at fixed Anderson disorder strength $U = 0.5$~eV. Currents are shown as functions of Fermi energy (a, d), $sp$ hopping $t_{sp}$ (b, e), and SOC strength $\lambda_{\rm SOC}$ (c, f). Dashed lines in each column mark the parameter values used in the corresponding panels of the other columns.

Panels~\ref{fig02}(a–c) show the charge current as a function of Fermi energy, $t_{sp}$, and $\lambda_{\rm SOC}$, highlighting the metallic regime where significant OHC and SHC can arise. The charge current serves as an indicator of the system’s conductivity and its capacity to support Hall currents.
Panel~\ref{fig02}(d) shows that both the OHC (blue) and SHC (red) are strongly dependent on $E_F$. The Hall currents peak near $-0.2$~eV (OHC) and 0.0~eV (SHC), close to $+0.1$~eV where the charge conductivity is maximal, indicating strong coupling between charge transport and Hall responses.

As shown in panel~\ref{fig02}(e), the OHC is highly sensitive to $t_{sp}$: it vanishes at $t_{sp} = 0$, increases with $t_{sp}$, and saturates for $t_{sp} > 1$. Since $t_{sp}$ controls the orbital texture~\cite{Go2018}, this confirms its relevance not only in clean systems but also under disorder. In contrast, the SHC shows weaker dependence on $t_{sp}$. While it vanishes at $t_{sp}=0$ and increases initially, it lacks a clear trend for $t_{sp} > 0.5$~eV and is primarily governed by SOC.

Panel \ref{fig02}(f)  contrasts the different impacts of SOC strength $\lambda_{\rm SOC}$ on OHC and SHC. As anticipated, the SHC  vanishes for $\lambda_{\rm SOC} = 0$, as the SOC is essential for generating spin currents.  In contrast, the OHC  is finite for $\lambda_{\rm SOC}=0$. This result generalizes the SOC-free emergence of the OHC, which was demonstrated for ideal disorder-free systems~\cite{Go2018}, to disordered systems. SHC and OHC behave differently as $\lambda_{\rm SOC}$ increases. SHC increases rapidly with $\lambda_{\rm SOC}$, though SHC decays for $\lambda_{\rm SOC}>0.4$ eV, where the charge current is suppressed [Fig.~\ref{fig02}(c)]. In contrast, OHC decreases almost monotonically with increasing $\lambda_{\rm SOC}$,  which is attributed primarily to the suppression of the charge current at high $\lambda_{\rm SOC}$ values. Notably, even with strong  SOC, the OHC remains significantly larger than SHC, underscoring the dominant role of orbital transport in the system.

\begin{figure}
\centering
\includegraphics[width=0.9\linewidth]{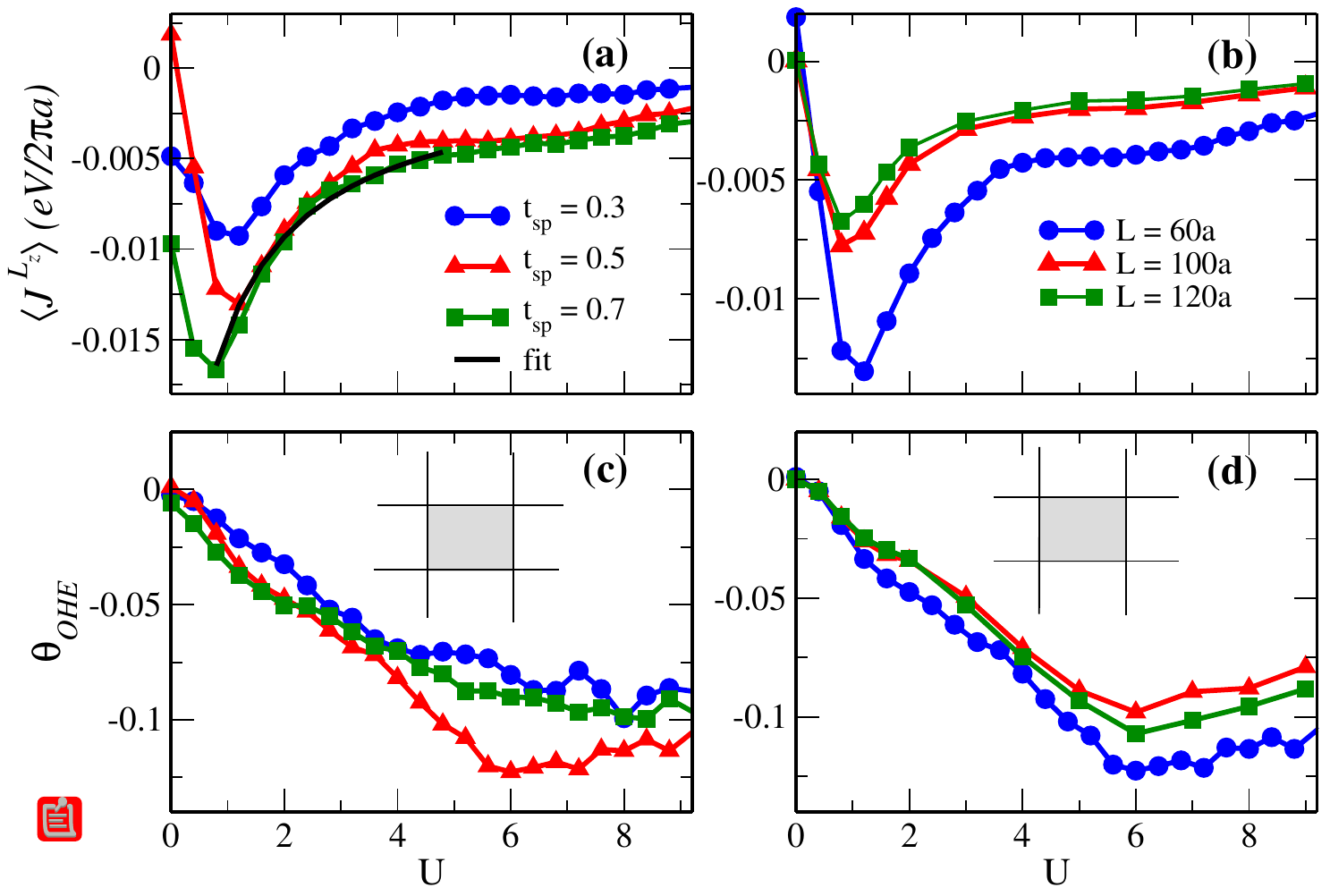}
\caption{(a,c) OHC density and (b,d) Orbital Hall angle for a square device, as illustrated in Fig.(\ref{fig01}.d), shown as a function of disorder strength $U$. The results are averaged over 300 disorder realizations, with $E_F = -0.7$~eV and no SOC($\lambda_{\rm SOC} = 0$). In (a,c), the device has a fixed size of $60a \times 60a$, and different $sp$ hopping values are considered: $t_{sp} = 0.3$, 0.5, and 0.7~eV. In (b,d), the results are shown for different device dimensions $L = 60a$, $100a$, and $120a$, with $t_{sp} = 0.5$~eV.}
\label{fig03}
\end{figure}

Figure~\ref{fig03} shows the OHC density $\langle {\cal J}^{L_z} \rangle = \langle I^{L_z} \rangle / L$ and orbital Hall angle as functions of Anderson disorder strength $U$, averaged over 300 disorder realizations for the square device in Fig.~\ref{fig01}(d). To simplify the analysis, we set $\lambda_{\rm SOC} = 0$, as realistic SOC values have negligible impact on our results.

Panels~(a) and (c) display results for fixed device size $60a \times 60a$ and varying $sp$ hopping $t_{sp} = 0.3$, 0.5, and 0.7~eV, showing how $t_{sp}$ influences both the magnitude and robustness of the orbital Hall response. Panels~(b) and (d) fix $t_{sp} = 0.5$~eV and vary the device size ($L = 60a$, $100a$, $120a$), demonstrating the scaling of OHC and angle with geometry. The behavior converges for $L \sim 100a$, indicating a saturation regime.

Figure~\ref{fig03}(a) clearly shows that disorder can significantly enhance the orbital Hall conductivity and even reverse the sign of the OHC, as observed in the system with $t_{sp}=0.5$ eV,  where the OHE is positive for $U=0$ and becomes negative for increasing values of $U$.

Further insights into the OHC behavior can be gained from its dependence on $U$. For intermediate disorder ($1~\mathrm{eV} \lesssim U \lesssim 4~\mathrm{eV}$), where charge transport is diffusive (see Fig.~S2 in SM), the OHC density follows $\langle {\cal J}^{L_z} \rangle \propto 1/(U + U_0)$ with $U_0 \sim 0.78$, as seen from the good fit for $t_{sp} = 0.7$~eV in Fig.~\ref{fig03}(a) (black solid line). This scaling is characteristic of an extrinsic OHE governed by skew scattering~\cite{Veneri2025}. This interpretation is reinforced by examining the orbital Hall angle $\Theta_{\rm OHE} = \langle I^{L_z} \rangle / \langle I \rangle$, which shows a linear dependence on $U$ [Fig.~\ref{fig03}(c)], also consistent with skew scattering~\cite{Veneri2025}.

For stronger disorder ($4~\mathrm{eV} \lesssim U \lesssim 8~\mathrm{eV}$), $|\langle {\cal J}^{L_z} \rangle|$ saturates at values comparable to the clean limit ($U = 0$). Together with the low-$U$ decay, this yields a clear rise-and-saturation behavior as a function of disorder.

A possible origin for the observed saturation involves the side-jump mechanism, where the extrinsic OHE becomes independent of $U$~\cite{Veneri2025}. This suggests a crossover from a skew-scattering–dominated regime to one dominated by side-jump processes, where $\langle {\cal J}^{L_z} \rangle$ no longer depends on impurity strength. At even larger $U$, the system enters a strongly disordered regime with suppressed charge transport (see Fig.~S2 in SM), indicating electronic localization. In this localized regime, the orbital Hall angle $\Theta_{\rm OHE}$ remains large or even increases [Fig.~\ref{fig03}(c)], consistent with a decoupling between orbital and charge currents.

 A similar decay-then-saturation behavior appears in the anomalous Hall effect (AHE) in ferromagnets~\cite{Nagaosa2010}. There, the Hall conductivity decreases with disorder in the weak-disorder (superclean) regime due to skew scattering, and saturates in the moderately dirty regime where the intrinsic contribution dominates. This analogy suggests a similar interpretation for Fig.~\ref{fig03}(a): at low $U$, skew scattering dominates and decreases with disorder; at higher $U$, the intrinsic mechanism dominates, remaining constant with increasing $U$.

 We now turn to a slightly different device configuration. Figure~\ref{fig04} examines the OHC and orbital Hall angle in a rectangular mesoscopic device with dimensions $60a \times L$, as a function of the device width $L$. In this geometry, the OHC $\langle I^{L_z} \rangle$ is generated within a square region of the device, specifically the area $L_0 \times L_0$. Beyond this region, the width of the sample is extended, creating an additional region of size $(L - L_0) \times L_0$ where the orbital current can propagate but is not actively generated. As $L$ increases, we expect a decay in the orbital current due to relaxation processes mediated by disorder. To examine this, we analyze the system under different disorder strengths $U$ and sample widths $L$.

Figure~\ref{fig04}(a) shows $\langle I^{L_z} \rangle$ as a function of $L$ for $U = 0.5$, 1.0, 2.0, and 3.0~eV, averaged over 600 disorder realizations, at a fixed Fermi energy $E_F = -0.7$~eV. The decay becomes more pronounced as $U$ increases, highlighting the role of disorder in orbital relaxation. For weak disorder ($U=0.5$ and $1.0$~eV), the decay is slow and can be described by either power law or exponential fits. At stronger disorder, the decay is clearly exponential.

Figure~\ref{fig04}(c) shows $\langle I^{L_z} \rangle$ on a logarithmic scale for $U=2.0$~eV, $U=3.0$~eV  and two Fermi energies ($E_F=-0.7$~eV and $E_F=-0.2$~eV). For both energies, the data are well fitted by an exponential decay, $\langle I^{L_z} \rangle \sim \exp(-L/\lambda_L)$, but the relaxation length $\lambda_L$ increases significantly when the Fermi energy shifts closer to the band degeneracy. This indicates that the degree of orbital degeneracy affects the suppression of relaxation processes.

Figure~\ref{fig04}(b) presents the orbital Hall angle $\Theta_{\rm OHE}$ as a function of $L$ for different disorder strengths at $E_F=-0.7$~eV. In this case, $\Theta_{\rm OHE}$ remains nearly constant with $L$, suggesting that the decay of the OAM currents follows the decay of the charge current.

Figure~\ref{fig04}(d) shows $\Theta_{\rm OHE}$ for the same disorder strength ($U=3.0$~eV) but comparing the two Fermi energies. Unlike the case at $E_F=-0.7$~eV, where $\Theta_{\rm OHE}$ is constant, at $E_F=-0.2$~eV the angle increases with distance. This implies that the orbital and charge currents do not decay at the same rate, and the relaxation mechanism depends sensitively on the details of the electronic structure.

The long orbital relaxation lengths observed in our disordered mesoscopic devices are on the same order as recent experimental estimates~\cite{Choi2023,Lyalin2023}, although those systems differ in both dimensionality and scale. We attribute this robustness to the suppression of orbital-flip scattering channels, a consequence of the system's two-dimensional character. In our model, the bands near the Fermi energy are primarily composed of $p_{x/y}$ orbitals, even though their degeneracy is lifted when $t_{sp}\neq0$ [Fig.~1(c)]. Orbital relaxation requires transitions between different angular momentum states, analogous to spin-flip processes in spin relaxation. These transitions are mediated by the orbital angular momentum ladder operators $L_\pm$, which couple $p_{x/y}$ to $p_z$. However, the absence of out-of-plane hopping and the lack of $t_{sp}$-induced hybridization with $p_z$ leave this orbital dynamically decoupled. As a result, scattering pathways capable of relaxing the $L_z$ component are strongly suppressed. While some orbital relaxation still occurs, it remains weak under these conditions and depends sensitively on the band structure near the Fermi level.

The exponential decay of the orbital current and the dependence of $\lambda_L$ and $\Theta_{\rm OHE}$ on Fermi energy and disorder strength suggest that multiple relaxation mechanisms coexist. In some regimes, an Elliott-Yafet-like process, where momentum scattering mediates orbital relaxation ($\lambda_L \propto \tau_p$), dominates, leading to similar scaling of orbital and charge currents. However, when the bands are more degenerate in orbital character, the suppression of phase-randomizing processes becomes more effective, and the decay of the orbital current can decouple from charge transport. In systems with explicit off-diagonal hopping or anisotropic disorder potentials that couple $p_x$ and $p_y$ directly, more efficient relaxation pathways would be expected.

\begin{figure}
\centering
\includegraphics[width=1.0\linewidth]{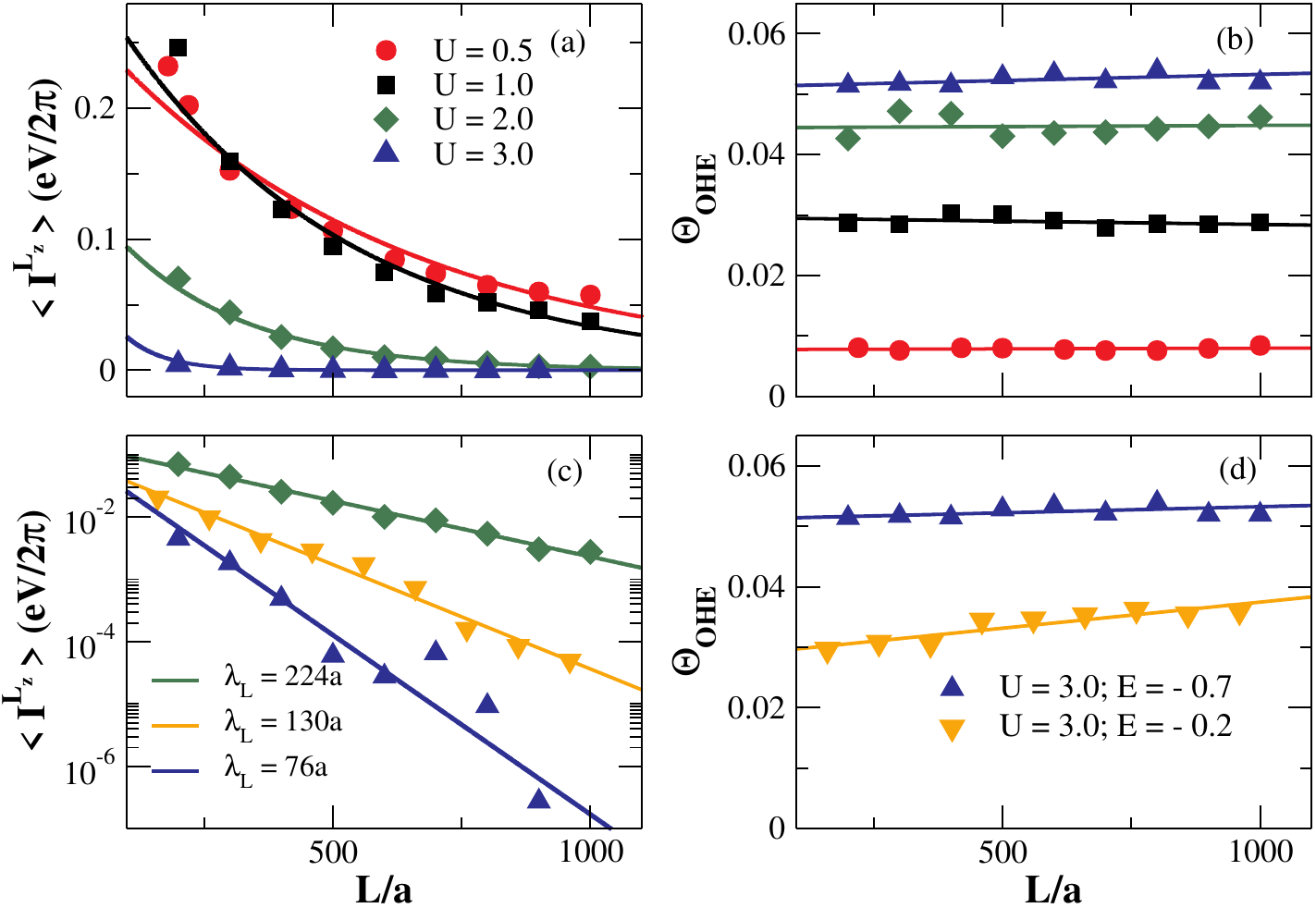}
\caption{(a) OHC, averaged over 600 disorder realizations (symbols), for a rectangular device with dimensions $60a \times L$, as illustrated in Fig.(\ref{fig01}.e). The current is shown as a function of the device width $L$, for parameters $t_{sp} = 0.5$, $E_F = -0.7$~eV, and various disorder strengths $U = 0.5$, 1.0, 2.0, and 3.0~eV. Panel (c) displays the data for $U = 2.0$ and $ 3.0$ eV at $E_F = -0.7$~eV, and $U = 3.0$ eV at $E_F = -0.2$~eV on a logarithmic scale, along with an exponential fit. (b,d) Orbital Hall angle as a function of the device width $L$, for the same parameters as in panels (a,c).}
\label{fig04}

\end{figure}

We numerically investigated the interplay between disorder and orbital transport in mesoscopic devices. In square geometries, disorder enhances the OHE via skew scattering, even reversing the orbital current sign—a clear signature of extrinsic mechanisms in the diffusive regime. In rectangular devices, the orbital current decays exponentially with width, revealing relaxation lengths that depend on disorder but remain remarkably long , highlighting the robustness of orbital transport. This resilience stems from a combination of symmetry protection and the nature of the disorder: in our strictly two-dimensional model, the $p_z$ orbital remains decoupled, which suppresses orbital-flip processes mediated by $L_\pm$ operators. Moreover, even relaxation within the $p_x/p_y$ subspace requires phase-randomizing scattering that effectively transforms one orbital character into the other. Since our scalar disorder is diagonal in the orbital basis, such processes are strongly suppressed or require higher-order virtual transitions involving the $s$ orbital. This mechanism is particularly inefficient when the bands are nearly degenerate, further extending the orbital relaxation length.

The near invariance of the orbital Hall angle with device width and its correlation with charge transport are consistent with an Elliott–Yafet-like relaxation mechanism, where momentum scattering mediates orbital decay. However, our results indicate that by tuning the Fermi energy and approaching stronger orbital degeneracy, the character of the relaxation can change, with the suppression of relaxation becoming more pronounced. This behavior emphasizes that the persistence of orbital coherence arises from the combined absence of direct $p_x$–$p_y$ coupling and the limited phase-randomizing channels in the scalar disorder model.

\begin{acknowledgments}
ALRB acknowledges financial support from Conselho Nacional de Desenvolvimento Cient\'{\i}fico e Tecnol\'ogico (CNPq, Grant 309457/2021) and INCT of Spintronics and Advanced Magnetic Nanostructures (INCT-SpinNanoMag), Grant No. CNPq 406836/2022-1. TGR acknowledges  FCT - Fundação para a Ciência e Tecnologia,  project reference numbers UIDB/04650/2020, 2023.11755.PEX ( with DOI identifier https://doi.org/10.54499/2023.11755.PEX) and 2022.07471.CEECIND/CP1718/CT0001
(with DOI identifier: 10.54499/2022.07471.CEECIND/CP1718/
CT0001) and acknowledges support from the EIC Pathfinder OPEN grant 101129641 “OBELIX”.
HWL acknowledges financial support from the National Research Foundation of Korea (NRF) grant funded by the Korean government (MSIT) (No. RS-2024-00410027). 
\end{acknowledgments}
\bibliography{ref}

\onecolumngrid
\renewcommand{\thefigure}{S\arabic{figure}}
\renewcommand{\theequation}{S\arabic{equation}}
\renewcommand{\thetable}{S\arabic{table}}

\setcounter{figure}{0}
\setcounter{equation}{0}
\setcounter{table}{0}

\newpage
\vskip 1cm
{\center \large \bf Supplementary material for ``Extrinsic Orbital Hall Effect and Orbital Relaxation in Mesoscopic Devices''}	
\vskip 1cm

To complement the analysis presented in the main text, we provide additional results highlighting the role of device size, disorder strength, and spin-orbit coupling (SOC) on the orbital Hall effect (OHE) and related transport properties. These results explore parameter regimes and trends that expand the conclusions drawn in the primary discussion.

Figure~\ref{figs0} illustrates the evolution of the band structure along the M–$\Gamma$–M path for increasing values of the $sp$ hopping parameter $t_{sp}$. Compared to Fig.~1(b–c) in the main text, these panels show that as $t_{sp}$ increases, the two bands initially split by hybridization become more dispersive. This enhanced dispersion reflects stronger mixing between the $s$ and $p_{x/y}$ orbitals. The $p_z$ orbital remains decoupled throughout as a result of the two-dimensional nature of the system. Notably, the increased hybridization alters the curvature and energy separation of the bands near the Fermi level, which directly impacts the intrinsic contribution to the orbital Hall effect discussed in the main text.

\begin{figure}[h]
\centering
\includegraphics[scale = 0.35]{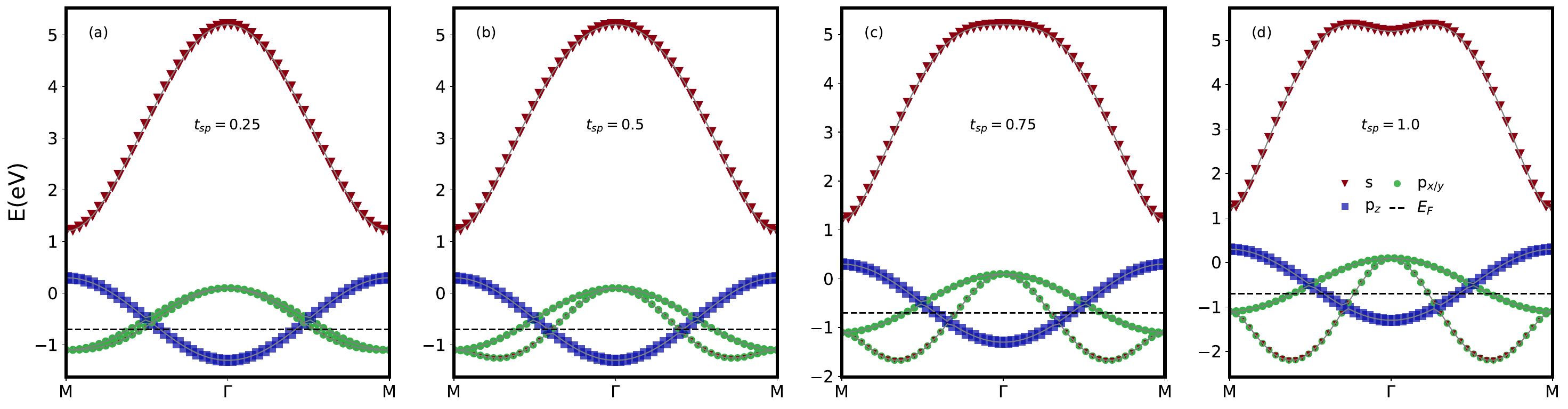}\caption{Energy bands and projected orbital character for the M–$\Gamma$–M path for increasing values of the $sp$ hopping parameter: (a) $t_{sp} = 0.25$ eV, (b) $t_{sp} = 0.5$ eV, (c) $t_{sp} = 0.75$ eV, and  (d) $t_{sp} = 1.0$ eV. Symbol sizes indicate orbital weights of $s$, $p_{x/y}$, and $p_z$ states. The dashed line marks the Fermi energy $E_F$.}\label{figs0}
\end{figure}

Figure \ref{figs1} presents the orbital Hall current density, $J^{L_z} = I^{L_z} / L$ (upper panels), and the orbital Hall angle, $\Theta_{\rm OHE}$ (lower panels), as functions of the lateral size $L$ of the square device. The aim is to analyze how the results scale with system size. Panels (a) and (c) display the results for different $sp$ hopping amplitudes, $t_{sp} = 0.5$, 0.7, and 1.0~eV, with fixed parameters $E_F = -0.7$~eV, $\lambda_{\rm SOC} = 0.0$, and $U = 0.5$eV. Panels (b) and (d) show the same quantities for varying disorder strengths, $U = 0.1$, 0.5, and 1.0eV, while keeping $t_{sp} = 0.5$~eV and $\lambda_{\rm SOC} = 0.0$.

Figure \ref{figs1} demonstrates that, apart from some oscillatory behavior, the results remain relatively stable for $L/a > 50$. The trends observed for $L/a \sim 60$ are qualitatively reliable, and the values converge for $L/a > 120$.

\begin{figure}[h]
\centering
\includegraphics[scale = 0.5]{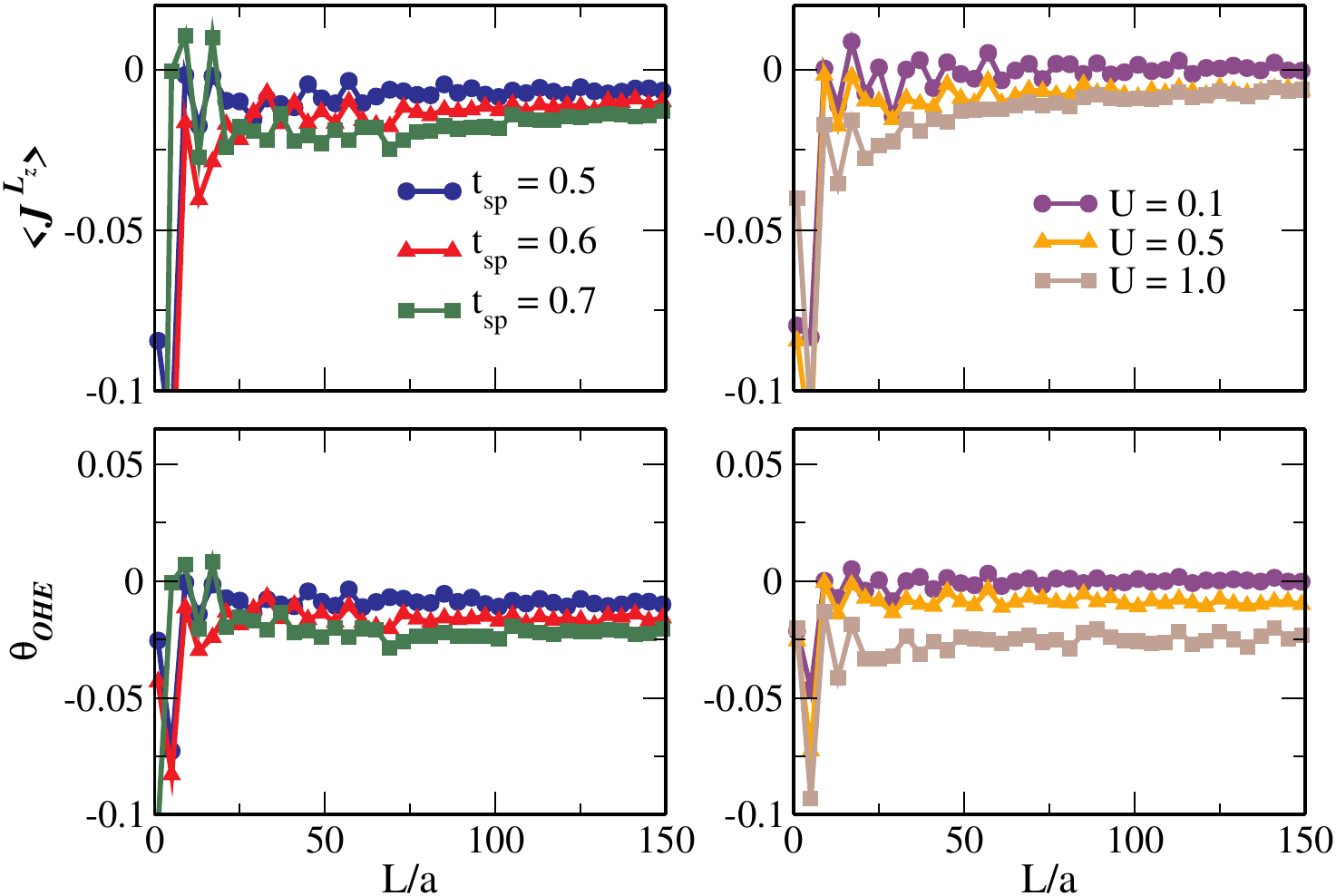}\caption{(a,b)
OH current density ($J^{L_z} = I^{L_z} / L$) and (c,d) OH angle, averaged over 100 realizations, as functions of the device lateral size $L$ of the square device. (a,c) Results for different $sp$ hopping values $t_{sp} = 0.5$, 0.7, and 1.0~eV, with fixed parameters $E = -0.7$~eV, $\lambda_{ \rm SOC} = 0.0$, and $U = 0.5$eV. (b,d) Results for different disorder strengths $U = 0.1$, 0.5, and 1.0eV, with fixed parameters $E = -0.7$~eV, $t_{sp} = 0.5$~eV, and  $\lambda_{\rm SOC} = 0.0$.}\label{figs1}
\end{figure}

Figure \ref{figs2} examines the interplay between disorder strength $U$, spin-orbit coupling (SOC), and transport properties, including the charge current, OHC, and SHC. The left panels correspond to a square device ($60a \times 60a$), while the right panels represent a rectangular device ($60a \times 80a$). For both geometries, the Fermi energy and $sp$ hopping are fixed at $E_F = -0.7$ eV and $t_{sp} = 0.5$ eV, while the spin-orbit coupling varies: $\lambda_{\rm SOC} = 0.0$, 0.1, and 0.2 eV. As expected, the charge current, shown in Fig.\ref{figs2}a and Fig.~\ref{figs2}b, decreases monotonically with increasing $U$, reflecting the suppression of carrier mobility caused by the disorder. The effect of SOC is also evident, with larger $\lambda_{\rm SOC}$ values leading to a slight reduction in the charge current.

The behavior of the OHC, depicted in Fig.\ref{figs2}c and Fig.\ref{figs2}d, is more complex. At low $U$, the OHC increases and reaches a peak at intermediate disorder strengths before decreasing at higher $U$. This non-monotonic trend aligns with the dominance of skew-scattering mechanisms at moderate disorder levels, as discussed in the main text. Interestingly, the OHC exhibits only a weak dependence on $\lambda_{\rm SOC}$, except at the minimum value of the OHC, where a slight dependence on SOC is observed.

In contrast, the SHC (Fig.\ref{figs2}e and Fig.\ref{figs2}f) strongly depends on $\lambda_{\rm SOC}$, vanishing entirely when $\lambda_{\rm SOC} = 0$, as expected. For strong $\lambda_{\rm SOC}$ and low $U$, the SHC increases with $U$ but remains much smaller in magnitude compared to the OHC. Unlike the OHC, the SHC does not exhibit a significant enhancement for intermediate $U$, indicating that disorder has a much weaker impact on spin transport. This reinforces the conclusion that orbital transport dominates over spin transport in these systems, with the OHC  being more sensitive to disorder effects.

\begin{figure}[h]
\centering
\includegraphics[scale = 0.5]{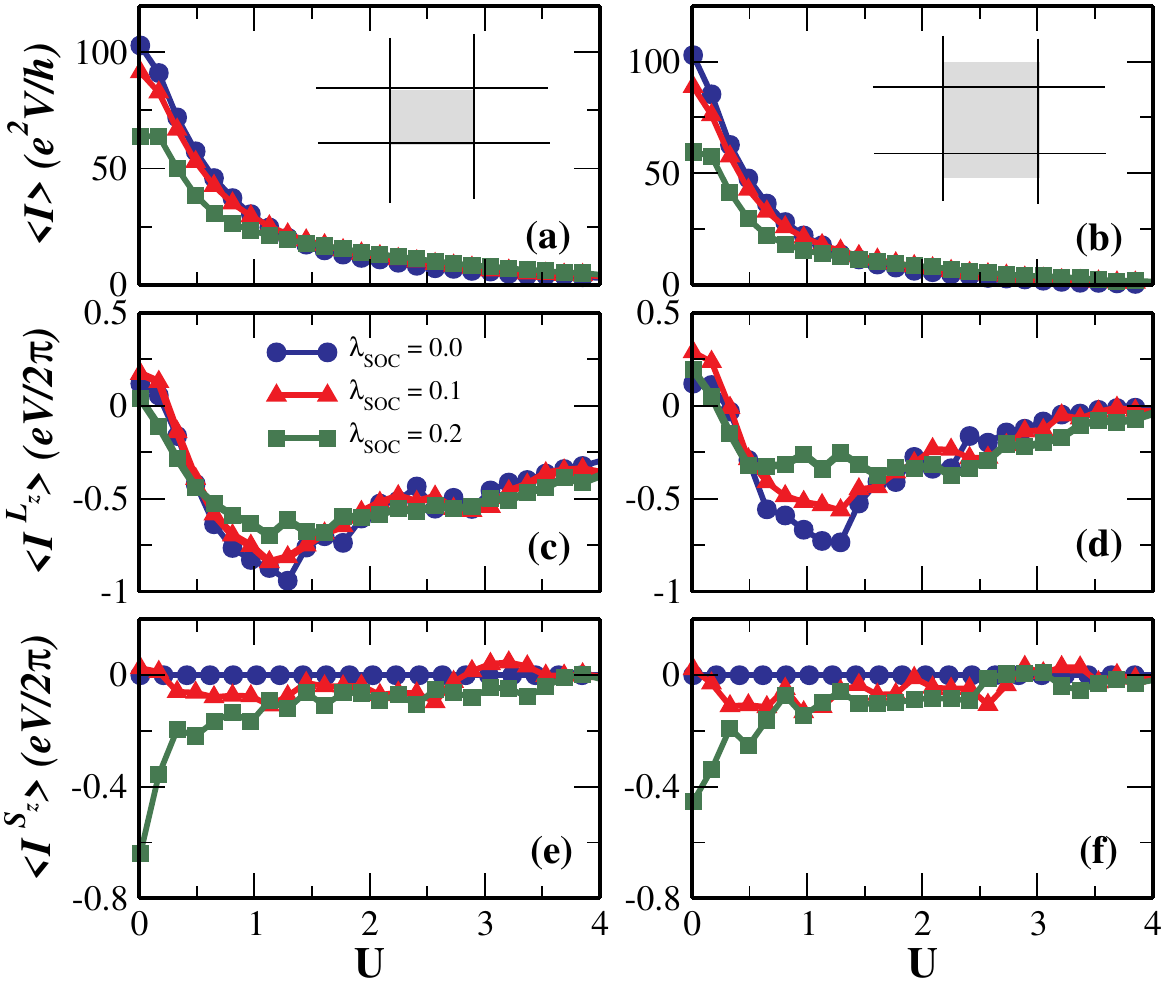}
\caption{(a,b) Charge current, (c,d) OHC, and (e,f) SHC, averaged over 100 realizations, as functions of disorder strength $U$ for different SOC values $\lambda_{\rm SOC} = 0.0$, 0.1, and 0.2~eV. The left panels correspond to a square device with dimensions $60a \times 60a$ [Fig.(\ref{fig01}.d)], while the right panels correspond to a rectangular device with dimensions $60a \times 80a$ [Fig.(\ref{fig01}.e)]. The parameters are fixed at $E = -0.7$~eV and $t_{sp} = 0.5$~eV.}\label{figs2}
\end{figure}

\end{document}